\documentclass[12pt,english,floatfix,superscriptaddress,aps,prd,preprint]{revtex4}
\usepackage[latin1]{inputenc}
\usepackage{amsmath}
\usepackage{amssymb}
\usepackage{amsbsy}
\usepackage{amsfonts}
\usepackage{amsopn}
\usepackage{amstext}
\usepackage{graphicx}
\usepackage{amssymb}
\usepackage{amsfonts}
\usepackage{amsmath}
\usepackage{graphicx}
\usepackage[english]{babel}
\usepackage{color}
\usepackage{esint}
\usepackage[dvips]{epsfig}
\usepackage[dvips]{graphicx}
\usepackage{units}
\usepackage{textcomp}

\usepackage[T1]{fontenc}
\usepackage{lmodern}
\usepackage{esint}
\usepackage{longtable}
\usepackage{babel} 
\usepackage{csquotes} 
\usepackage{color} 
\usepackage{textcomp}

\usepackage{hyperref}
\usepackage{slashed}

\usepackage{siunitx} % adicionado por Victor
\usepackage{graphicx} % adicionado por Victor 2

\newcommand{\be}{\begin{equation}}
\newcommand{\ee}{\end{equation}}
\newcommand{\ben}{\begin{eqnarray}}
\newcommand{\een}{\end{eqnarray}}

\begin{document}

\title{Softly higher-derivative massive gravity}

\author{C. A. S. Almeida}
\email{carlos@fisica.ufc.br}
\affiliation{Universidade Federal do Cear\'a (UFC), Departamento de F\'isica,\\ Campus do Pici, Fortaleza - CE, C.P. 6030, 60455-760 - Brazil.}

%%%%%%%%%%%%%%%%%%%%%%%%%%%%%%%%%%%%%%%%%%%%%%%%%%%%%%%%%%%%%%%%%%%%%%%%%%%%%%%%%%%%%%%%%%%%%%%%%%
%%%%%%%%%%%%%%%%%%%%%%%%%%%%%%%%%%%%%%%%%%%%%%%%%%%%%%%%%%%%%%%%%%%%%%

\author{W. T. Cruz}
\email{wilamicruz@gmail.com}
\affiliation{Instituto Federal de Educa\c{c}\~ao, Ci\^encia e Tecnologia do Cear\'a (IFCE),
	Campus Juazeiro do Norte, 63040-000 Juazeiro do Norte - CE - Brazil }

%%%%%%%%%%%%%%%%%%%%%%%%%%%%%%%%%%%%%%%%%%%%%%%%%%%%%%%%%%%%%%%%%%%%%%%

\author{R. V. Maluf}
\email{r.v.maluf@fisica.ufc.br}
\affiliation{Universidade Federal do Cear\'a (UFC), Departamento de F\'isica,\\ Campus do Pici, Fortaleza - CE, C.P. 6030, 60455-760 - Brazil.}

%%%%%%%%%%%%%%%%%%%%%%%%%%%%%%%%%%%%%%%%%%%%%%%%%%%%%%%%%%%%%%%%%%%%%%

\author{A. Yu. Petrov}
\email{petrov@fisica.ufpb.br}
\affiliation{Departamento de F\'{\i}sica, Universidade Federal da Para\'{\i}ba\\
 Caixa Postal 5008, 58051-970, Jo\~ao Pessoa, PB, Brazil}

\author{P. Porfirio}
\email{pporfirio@fisica.ufpb.br}
\affiliation{Departamento de F\'{\i}sica, Universidade Federal da Para\'{\i}ba\\
	Caixa Postal 5008, 58051-970, Jo\~ao Pessoa, PB, Brazil}

\date{\today}

\begin{abstract}
In this work, we study the higher-derivative massive gravity in $D$-dimensional space-time with the
mass term arisen due to a 2-brane embedded in a $4D$ Minkowski spacetime. We consider the effect
of a resonance mass term from the DGP braneworld model for the higher-derivative massive
gravity. Moreover, we add the gravitational Chern-Simons and Ricci-Cotton terms to this
model and evaluate the graviton propagator using a basis of Barnes-Rivers spin projectors.
Using the obtained propagator, we proceed with discussing the consistency of the
model, writing the dispersion relations, and analyzing causality and unitarity.
 \end{abstract}

%\keywords{*****}
\maketitle

%%%%%%%%%%%%%%%%%%%%%%%%%%%%%%%%%%%%%%%%%%%%%%%%%%%%%%%%%%%%%%%%%%%%%%%%%%%%%%%%%%
\section{Introduction}

Over the last years, theories of massive gravity has attracted great attention. Essentially, the idea of possibility of mass for gravitons is naturally related with the facts that, first, the gravity, similarly to non-abelian Yang-Mills theory, displays self-interaction, second, the mass is a natural measure of interaction of any particles or fields with gravitational field. First steps in development of the massive gravity concept were done already by Fierz and Pauli \cite{Fierz:1939ix} who considered a theory of massive spin-2 field on a flat background.

Further,  the idea of massive gravitons has been abandoned for a long time. Besides of observations which were not confirming non-zero mass of graviton, an essential reason of this was related with a discovery of the Boulware-Deser ghost \cite{Boulware:1972yco} which makes problematic an introduction of interaction of massive gravity with a matter. However, recently interest to massive gravity returned, first of all, due to evidence of dark energy \cite{SupernovaSearchTeam:1998fmf} whose explanation requires a modification of gravity theory, so, various extensions of gravity must be tested.

There are various ways to introduce massive gravity now. In three dimensions, the most interesting theory in this context is the Bergshoeff-Hohm-Townsend (BHT) gravity \cite{Bergshoeff:2009hq}, whose generalization to four-dimensional space-time has been proposed in \cite{Bergshoeff:2012ud}. Another important manner to construct the massive gravity is presented by de Rham-Gabadadze-Tolley (dRGT) theory \cite{deRham:2010kj} representing itself as a specific nonlinear extension of the Fierz-Pauli gravity in a manner allowing to rule out the Boulware-Deser ghost. One more non-linear ghost-free massive gravity model has been proposed in \cite{Hassam1,Hassam2}. Among these ways, an important role is played by Dvali-Gabadadze-Porrati (DGP) gravity \cite{Dvali:2000hr} whose importance consists in the fact that it essentially includes an extra dimension. Namely this theory will be the main object of study in this paper. To be more precise, we consider a $D$-dimensional gravity model whose action is composed of the usual Einstein-Hilbert term added by quadratic terms in the curvature plus a Fierz-Pauli-like mass term, with a momentum dependent operator (for the braneworld approach in higher-derivative gravity, see also \cite{Od1,Od2}). The latter has its origins justified in the context of extra dimensions in the DGP theory. Moreover, we add the topological Chern-Simons term and its higher-derivative version, the Ricci-Cotton term, to this model in three dimensions. Within this study, our aim consists in studying the structure of the graviton propagator and the modifications in the gravitational potential arising due to different mass terms.

The paper is organized as follows. In Sec.~\ref{sec2} we briefly revise the DGP scenario for lower dimensions and linearize the $D$-dimensional action up to quadratic terms. Afterwards, we invert the kernel of the action to find the graviton propagator. Also, we study the structure of the propagators to identify the particle content of the theory. In Sec.~\ref{sec3}, since the `resonance massive graviton' can be generalized, we address the issue of graviton propagator with generalized mass term to find several scenarios with distinct interparticle potentials. In Sec. ~\ref{sec4} we consider the more general situation where the gravity action involves Ricci-Cotton and Chern-Simons terms. Finally in Sec.~\ref{conclu} we make our final considerations. Within this paper, we employ the following conventions and definitions:
the flat metric is $\eta_{\mu\nu}=\mbox{Diag}(-1,+1,+1,\cdots,+1)$, the Riemann tensor is $R^{\alpha}_{\ \ \beta\gamma\delta}=\partial_{\delta}\Gamma^{\alpha}_{\beta\gamma}-\partial_{\gamma}\Gamma^{\alpha}_{\beta\delta}+\Gamma^{\alpha}_{\gamma\lambda} \Gamma^{\lambda}_{\beta\sigma} - \Gamma^{\alpha}_{\delta\lambda} \Gamma^{\lambda}_{\gamma\beta} $  and the connection is given by the Cristoffel symbols, $\Gamma^{\mu}_{\alpha\beta}=\frac{1}{2}g^{\mu\lambda}\left(-\partial_{\lambda}g_{\alpha\beta}+\partial_{\alpha}g_{\lambda\beta}+\partial_{\beta}g_{\alpha\lambda}\right)$.

%%%%%%%%%%%%%%%%%%%%%%%%%%%%%%%%%%%%%%%%%%%%%%%%%%%%%%%%%%%%%%%%%%%%%%%%%%%%%%%%%%%%%%%%%%%%%%%%%%

\section{Propagator in a softly higher-derivative massive gravity }
\label{sec2}

In this section, we revisit the massive gravity theory induced by extra dimensions whose mass term for the graviton is a function of the momentum. This model is known as the Dvali-Gabadadze-Porrati (DGP) braneworld model \cite{Dvali:2000hr}. The DGP model contains a 3-brane setup embedded in a five-dimensional bulk spacetime. We wish to pursue the possibility of constructing a well-behaved higher-derivative massive gravity with a momentum-dependent mass term by studying the structure of the graviton propagator in $D$ dimensions.

The action which describes the gravity in the DGP scenario can be split into two parts, namely
\begin{equation}
S=S_{(5)}+S_{(4)},
\label{ac1}
\end{equation}
with
\begin{equation}
S_{(5)}=\frac{M^{3}_{5}}{2}\int d^{5}x\sqrt{-G}R^{(5)},\label{S5}
\end{equation}
describing the Einstein-Hilbert action in a five-dimensional bulk spacetime. It must be realized that $R^{(5)}$ is the five-dimensional Ricci scalar constructed from the bulk metric. The second part in the action (\ref{ac1}) is 
\begin{equation}
S_{(4)}=\frac{M^{2}_{4}}{2}\int d^{4}x \sqrt{-g}R+S_{M}(g_{\mu\nu},\psi).
\end{equation}
It includes the usual Einstein-Hilbert action in four dimensions plus the contributions stemming from the matter sources, which are restricted to couple with the four-dimensional metric only. Here, we pick the metric $G_{AB}$ to denote the five-dimensional bulk spacetime while $g_{\mu\nu}$ denotes the induced metric on the four-dimensional brane. Furthermore, we choose the following convention, namely, capital letters label bulk coordinates ($A=0,...,4$) and small letters label ``parallel'' coordinates ($m=0,..,3$) to the brane. One usually split the bulk coordinates into two components: $x^{A}=(x^m, y)$ and, thus, $y$ label the extra dimension coordinate which, in turn, is ``transverse'' to the brane, see for example \cite{Cvetic:2020axz}. We also assume the following standard boundary condition: $G_{AB}(x^m,y)\big|_{y=0}\equiv g_{AB}(x^m)$ in which entails that the $3$-brane is placed at $y=0$.

To examine the particle content of the DGP model, we can expand the $5D$ metric into the form  $G_{AB}(x^m,y)=\eta_{AB}+h_{AB}(x^m,y)$, where $h_{AB}(x^m,y)$ are the metric fluctuations around the flat spacetime. As a consequence, one can define the $4D$ metric fluctuations as $h_{AB}(x^m,y)\big|_{y=0}\equiv h_{AB}(x^m)$.
To proceed with the dimensional reduction, we must integrate over the bulk coordinate and thus obtain an effective $4D$ action. Following the standard procedure outlined in Refs. \cite{Luty:2003vm,Hao:2014tsa,Bazeia:2014xfa}, it is convenient to adopt the ADM-like variables adapted to the DGP case (see \cite{Hinterbichler:2011tt} for details), and it can be shown that the action induced for (\ref{S5}) in $4D$ takes the following form
\begin{equation}
S_{(5)}+S_{gf}=\frac{M^{2}_{4}}{2}\int d^{4}x\left(-\frac{1}{2}mh_{mn}\sqrt{-\Box} h^{mn}+\frac{1}{2}mh\sqrt{-\Box} h\right),\label{S5to4}
\end{equation} 
where the gauge-fixing term is based on de Donder gauge, being equal to
\begin{equation}
S_{gf}=-\frac{M_{5}^{2}}{2}\int d^{5}x\left(\partial^{A}h_{AB}-\frac{1}{2}\partial_{B}h\right)^{2},
\end{equation}
and $m\equiv\frac{M_{5}^{3}}{M_{4}^{2}}$, where $M_{5}$ is the mass scale on the bulk and $M_{4}$ is on the brane is known as the DGP or crossover scale. It is worth noting that the induced action (\ref{S5to4}) presents the Fierz-Pauli form with an operator-dependent mass term $m\sqrt{-\Box}$. This resonance mass (or \textit{soft} mass) \cite{Gabadadze:2003ck} can be generalized assuming an arbitrary dependence on the d'Alembert operator, $m\sqrt{-\Box}\rightarrow m^{2}(\Box)$. If we take a Taylor expansion for large distances (or small momenta), the dominant term must take the form \cite{Dvali:2007kt}:
\begin{equation}
m^{2}(\Box)=L^{2(\alpha-1)}(-\Box)^{\alpha},\label{massfunction}
\end{equation}
with $L$ representing a length scale parameter and $\alpha$ being a constant. Modifying Newtonian gravity at large scales requires the mass term to dominate over two-derivative kinetic terms, so we should have $\alpha<1$. 
As a consequence, our theory effectively becomes a nonlocal one since $\alpha$ is fractional. However, unlike the conventional nonlocal gravity theories (see f.e. \cite{Modesto} and references therein), the new term, first, involves a fractional degree of the d'Alembertian operator while in the usual cases only integer degrees of $\Box$ are present, second, dominates in the infrared limit instead of the ultraviolet one as occurs f.e. in \cite{Modesto}, thus being similar to the $\ln \Box$ terms arising within quantum corrections in many situations.
On the other hand, there is the requirement for the spectral
function to be positive definite, so that there are no ghosts. So we have a lower bound $\alpha>0$ \cite{Dvali:2006su}. The standard DGP model corresponds to $\alpha=1/2$ and $L=1/m$.  A less explored case in the literature corresponds to the limit $\alpha\rightarrow 1$ \cite{Bazeia:2014xfa}.

After this brief review of the DGP model, let us now consider a higher-derivative gravity model added to a Pauli-Fierz-like \textit{softly} massive term in $D$ dimensions and examine its effect on the graviton propagator.

For this purpose, we assume the the following action up to quadratic order in the curvature:
\begin{equation}
S=\int d^D x\left[\sqrt{-g}\frac{2}{\kappa^{2}}\left\{R+\frac{\beta}{2}R^{2}+\frac{\gamma}{2}R^{2}_{\mu\nu}\right\}-\frac{1}{2}\left(h_{\mu\nu}m^{2}(\Box)h^{\mu\nu}-h m^{2}(\Box) h \right)\right],\label{action1}
\end{equation}
where $\kappa^{2}$ is a suitable constant with mass dimension $[\kappa^{2}]=M^{2-D}$ (in natural units) which in four dimensions is equal to $32\pi G_{N}$, such that $G_{N}$ is the Newtonian gravitational
constant. Also, $\beta$ and $\gamma$ are constants with mass dimension $M^{-2}$, and $m^{2}(\Box)$ is the mass function defined in (\ref{massfunction}). 

As before, we will take the weak field approximation by decomposing the metric as $g_{\mu\nu}=\eta_{\mu\nu}+\kappa h_{\mu\nu}$ and keep only quadratic fluctuations in the action \eqref{action1}. This procedure leads to the following expression:
\begin{eqnarray}
S=\int d^D x \!\!\!\! && \left[\frac{1}{2}h^{\mu\nu}\Box h_{\mu\nu}-\frac{1}{2}h \Box h+h^{\mu\nu}\partial_{\mu}\partial_{\nu}h-h^{\mu\nu}\partial_{\mu}\partial_{\sigma} h^{\sigma}_{\  \nu}
-\frac{1}{2}\left(h^{\mu\nu}m^{2}(\Box)h_{\mu\nu}-h m^{2}(\Box) h \right)\right.\nonumber\\
&&\left.+\beta\left(h^{\mu\nu}\partial_{\mu}\partial_{\nu}\partial_{\alpha}\partial_{\beta}h^{\alpha\beta}
-2h^{\mu\nu}\Box\partial_{\mu}\partial_{\nu}h
+h\Box^{2}h\right)+\frac{\gamma}{4}\left(h^{\mu\nu}\Box^{2}h_{\mu\nu}+h\Box^{2}h\right.\right.\nonumber\\
&&\left.\left. -2h^{\mu\nu}\Box\partial_{\mu}\partial_{\lambda}h^{\lambda}_{\ \nu}+2h^{\mu\nu}\partial_{\mu}\partial_{\nu}\partial_{\lambda}
\partial_{\sigma}h^{\lambda\sigma}-2h\Box\partial_{\mu}\partial_{\nu}h^{\mu\nu}\right)\right].\label{action2}
\end{eqnarray}

It is convenient to rewrite the action \eqref{action2} in terms of a set of spin projection operators in the space of symmetric rank-two tensors. The complete set of spin projection operators in $D$ dimensions, $P^{(2)}$, $P^{(1)}$, $P^{(0-s)}$, $P^{(0-w)}$, $P^{(0-sw)}$, and $P^{(0-ws)}$ are defined as follows \cite{Nakasone:2009vt}:
\begin{eqnarray}
P^{(2)}_{\mu\nu, \rho\sigma}&=&\frac{1}{2}(\theta_{\mu\rho}\theta_{\nu\sigma}+\theta_{\mu\sigma}\theta_{\nu\rho})-\frac{1}{D-1}\theta_{\mu\nu}
\theta_{\rho\sigma},\nonumber\\
P^{(1)}_{\mu\nu, \rho\sigma}&=&\frac{1}{2}(\theta_{\mu\rho}\omega_{\nu\sigma}+\theta_{\mu\sigma}\omega_{\nu\rho}+\theta_{\nu\rho}
\omega_{\mu\sigma}+\theta_{\nu\sigma}\omega_{\mu\rho}),\nonumber\\
P^{(0,s)}_{\mu\nu, \rho\sigma}&=&\frac{1}{D-1}\theta_{\mu\nu}\theta_{\rho\sigma},\;\; P^{(0,w)}_{\mu\nu, \rho\sigma}=\omega_{\mu\nu}\omega_{\rho\sigma},\nonumber\\
P^{(0,sw)}_{\mu\nu, \rho\sigma}&=&\frac{1}{\sqrt{D-1}}\theta_{\mu\nu}\omega_{\rho\sigma},\;\;\;\;\; P^{(0,ws)}_{\mu\nu, \rho\sigma}=\frac{1}{\sqrt{D-1}}\omega_{\mu\nu}\theta_{\rho\sigma},\label{spinProj1}
\end{eqnarray} where the transverse and the longitudinal operators ($\theta_{\mu\nu}$ and $\omega_{\mu\nu}$) are given by
\be \theta_{\mu\nu}=\eta_{\mu\nu}-\frac{1}{\Box}\partial_{\mu}\partial_{\nu}=\eta_{\mu\nu}-\omega_{\mu\nu},\;\;\;\;\; \omega_{\mu\nu}=\frac{1}{\Box}\partial_{\mu}\partial_{\nu},\ee
and satisfy the following relationships:
\be \theta_{\mu\rho}\theta^{\rho}_{\;\;\nu}=\theta_{\mu\nu},\;\;\;\;\;\omega_{\mu\rho}\omega^{\rho}_{\;\;\nu}=\omega_{\mu\nu},\;\;\;\;\;\theta_{\mu\rho}\omega^{\rho}_{\;\;\nu}=0.\ee

The multiplicative table for the projecting operators can be assembled through the orthogonality relations:
\begin{eqnarray}
P_{\mu\nu,\rho\sigma}^{(i,a)}P_{\phantom{(j,b)\rho\sigma,}\lambda\tau}^{(j,b)\rho\sigma,}&=&\delta^{ij}\delta^{ab}P_{\mu\nu,\lambda\tau}^{(i,a)},\nonumber\\
P_{\mu\nu,\rho\sigma}^{(i,ab)}P_{\phantom{(j,b)\rho\sigma,,}\lambda\tau}^{(j,cd)\rho\sigma,}&=&\delta^{ij}\delta^{bc}P_{\mu\nu,\lambda\tau}^{(i,a)},\nonumber\\
P_{\mu\nu,\rho\sigma}^{(i,a)}P_{\phantom{(j,b)\rho\sigma,,}\lambda\tau}^{(j,bc)\rho\sigma,}&=&\delta^{ij}\delta^{ab}P_{\mu\nu,\lambda\tau}^{(i,ac)},\nonumber\\
P_{\mu\nu,\rho\sigma}^{(i,ab)}P_{\phantom{(j,b)\rho\sigma,,}\lambda\tau}^{(j,c)\rho\sigma,}&=&\delta^{ij}\delta^{bc}P_{\mu\nu,\lambda\tau}^{(i,ac)},
\label{orthog}\end{eqnarray}with $i,j=0,1,2,$ and $a,b,c,d=s,w$ and the tensorial identity:
\begin{equation}
\left[P^{(2)}+P^{(1)}+P^{(0,s)}+P^{(0,w)}\right]_{\mu\nu,\rho\sigma}=\frac{1}{2}\left(\eta_{\mu\rho}\eta_{\nu\sigma}+\eta_{\mu\sigma}\eta_{\nu\rho}\right)\equiv \mathcal{I}_{\mu\nu,\rho\sigma}.
\label{identity}\end{equation}

Using the relations \eqref{spinProj1}, \eqref{orthog} and \eqref{identity}, the action \eqref{action2} can be rewritten as
\be S=\int d^{D}x\frac{1}{2}h^{\mu\nu}\mathcal{O}_{\mu\nu,\alpha\beta} h^{\alpha\beta},\ee where the operator $\mathcal{O}_{\mu\nu,\alpha\beta}$ is identified as
\begin{eqnarray}
\mathcal{O}_{\mu\nu,\alpha\beta}&=&\left[\frac{\gamma}{2}\Box^{2}+\Box -m^{2}(\Box)\right]P^{(2)}_{\mu\nu,\alpha\beta}-m^{2}(\Box) P^{(1)}_{\mu\nu,\alpha\beta}\nonumber\\
&&+\left[\left(2\beta(D-1)+\frac{\gamma}{2}D\right)\Box^{2}-(D-2)\left(\Box-m^{2}(\Box)\right)\right]P^{(0,s)}_{\mu\nu,\alpha\beta}\nonumber\\
&&+\sqrt{D-1}m^{2}(\Box)(P^{(0,sw)}+P^{(0,ws)})_{\mu\nu,\alpha\beta}.
\end{eqnarray}

Taking into account that $\mathcal{O}\mathcal{O}^{-1} =\mathcal{I}$, we can find the graviton propagator inverting each spin block
\begin{eqnarray}\label{invO}
\mathcal{O}^{-1}_{\mu\nu,\alpha\beta}	&=&	\left[\frac{1}{\frac{\gamma}{2}\Box^{2}+\Box-m^{2}(\Box)}\right]P^{(2)}_{\mu\nu,\alpha\beta}-\left[\frac{1}{m^{2}(\Box)}\right]P^{(1)}_{\mu\nu,\alpha\beta}\nonumber\\
&+&	\left[-\frac{ \left(4 \beta  (D-1)+\gamma  D\right)\Box^2+(4-2 D) \left(\Box-m^2(\Box)\right)}{2 (D-1) (m^2(\Box))^2}\right]P^{(0,w)}_{\mu\nu,\alpha\beta}\nonumber\\
    &+&\left[\frac{1}{\sqrt{D-1}m^{2}(\Box)}\right](P^{(0,sw)}+P^{(0,ws)})_{\mu\nu,\alpha\beta}.
\end{eqnarray}

Let us discuss unitarity and causality of this propagator, i.e. check presence of ghosts (negative residues) and tachyons (poles $\mu^2$ of the propagator corresponding to space-like momenta). The expression of the propagator (\ref{invO}) reveals that only the spin-projector $P^{(2)}$ will make a non-zero contribution to the tree-level current-current amplitude since $k_{\mu}T^{\mu\nu}=0$, where the external sources are assumed to be conserved with transverse stress-energy tensor. Thus, the possible massive poles associated with the spin 2 sector can be read off from the function
\begin{equation}
I(\gamma,\alpha)\equiv\frac{\gamma}{2}\Box^{2}+\Box-(-1)^{\alpha}L^{2(\alpha-1)}\Box^{\alpha},
\end{equation} where we take the expression (\ref{massfunction}) for the mass term $m^2(\Box)$. The pole structure indicates that the absence or not of ghosts and tachyons modes will depend on the particular choice of parameters. In fact, the necessary condition for unitarity of our theory consists in impossibility of expansion of the denominator $\frac{\gamma}{2}\Box^2+\Box-m^2(\Box)$ cannot be expanded into a product of two primitive multipliers, that is, $(\Box-\omega_{+})(\Box-\omega_{-})$, so that, as a consequence, the factor $\frac{1}{\frac{\gamma}{2}\Box^2+\Box-m^2(\Box)}$ characterizing our propagator cannot be expanded in a sum of two primitive multipliers, therefore, there is no different-sign residues and hence no ghosts in our theory. Let us verify this condition explicitly in some cases. For $\gamma > 0$ and $\alpha=0$, the equation $I=0$ has two real solutions:
\begin{equation}
I(\gamma,0)=\frac{\gamma}{2}(\Box-\omega_{+})(\Box-\omega_{-}),
\end{equation}where $\omega_{\pm}\equiv\frac{1}{\gamma}\left(-1\pm\sqrt{1+2 \frac{\gamma}{L^{2}}}\right)$, such that $\omega_{+}>0$ and $\omega_{-}<0$. It is easy to see that we can rewrite $1/I(\gamma,0)$ as
\begin{equation}
\frac{1}{I(\gamma,0)}=\frac{1}{\sqrt{1+ 2\frac{\gamma}{L^{2}}}}\left(\frac{1}{\Box-\omega_{+}}-\frac{1}{\Box-\omega_{-}}\right),
\end{equation}which shows that there are two massive modes. One mode is unitary with positive mass $\sqrt{\omega_{+}}$ and positive norm while the other is the ghost with tachyonic mass $i\sqrt{|\omega_{-}}|$ and negative norm. This corresponds to the case of the massive gravity theory studied in Ref. \cite{Nakasone:2009vt}. The next case we analyze corresponds to the standard DGP model where $\gamma=0$, $\alpha=1/2$ and $L=1/m$. The pole in this case can be read off from
\begin{equation}
I(0,1/2)=\Box-m\sqrt{-\Box},
\end{equation}and unlike the previous case it cannot be expanded in the form of the product of real roots. It is well known that the DGP theory's propagator has a well-defined spectral representation, and its spectrum contains a continuum of ordinary (unitary and causal) gravitons, with masses ranging from 0 to infinity \cite{Hinterbichler:2011tt}. Finally, we will analyze the least explored case in the literature when $\alpha\rightarrow 1$. The pole is described by
\begin{equation}
I(\gamma,1)=\frac{\gamma}{2}\Box(\Box+\frac{4}{\gamma}), 
\end{equation}and the corresponding propagator is
\begin{equation}
\frac{1}{I(\gamma,1)}=\frac{1}{2}\left(\frac{1}{\Box}-\frac{1}{\Box+\frac{4}{\gamma}}\right).
\end{equation}This structure indicates the presence of two poles, one massless unitary mode and a massive ghost mode which is causal for $\gamma<0$ and tachyonic for $\gamma>0$. In conclusion, the non-unitary and tachyonic modes in general can be present in our theory.

%%%%%%%%%%%%%%%%%%%%%%%%%%%%%%%%%%%%%%%%%%%%%%%%%%%%%%%%%%%%%%%%%%%%%%%%%%%%%%%%%%
\section{Interparticle potential}
\label{sec3}

Now we intend to investigate the modifications of the Newtonian potential due to the presence of the soft mass term coexisting with higher-derivative gravity. To this aim, we calculate the effective nonrelativistic potential of interaction between two identical massive spin-zero bosons via a graviton exchange.

The free graviton propagator satisfies the Green's equation, given by
\begin{equation}
\mathcal{O}^{\mu\nu,}_{\ \ \lambda\sigma}D_{F}^{\lambda\sigma,\alpha\beta}(x-y)=i\mathcal{I}^{\mu\nu,\alpha\beta}\delta^{(D)}(x-y),
\end{equation}
where the inverse of the wave operator $\mathcal{O}$ is expressed in (\ref{invO}). Thus, the Feynman propagator takes the form in the momentum space
\begin{equation}
D_{F}^{\mu\nu,\alpha\beta}(p)=\frac{i}{\frac{\gamma}{2}p^{4}-p^{2}-m^{2}(-p^{2})}\times \left[\frac{1}{2}(\eta^{\mu\alpha}\eta^{\nu\beta}+\eta^{\mu\beta}\eta^{\nu\alpha})-\frac{1}{D-1}\eta^{\mu\nu}\eta^{\alpha\beta}\right],\label{SoftlyProp}
\end{equation}
in which only the contribution of the projection operator $P^{(2)}$ is taken into account due to the current conservation. Note that for $D = 4$, $\gamma=0$ and $m^{2}(-p^{2})=m\sqrt{p^{2}}$ (choosing $\alpha=1/2$ and $L^{-1}\equiv m$), we recovered the graviton propagator from the original DGP model \cite{Hinterbichler:2011tt}.

The next step is to choose an external source for gravity. Let us consider the matter content being represented by the following action,
\begin{equation}
S_{\mbox{matter}}=\int d^{D}x\sqrt{-g}\left[-\frac{1}{2}g^{\mu\nu}\partial_{\mu}\phi\partial_{\nu}\phi-\frac{1}{2}M^{2}\phi^{2}\right],
\end{equation} 
which describes the dynamics of a single real scalar field in curved spacetime. Carrying out the weak field expansion up to first order in $h_{\mu\nu}$, we obtain the following Lagrangian:
\begin{eqnarray}
\mathcal{L}_{\mbox{matter}} & \thickapprox & -\frac{1}{2}\eta^{\mu\nu}\partial_{\mu}\phi
\partial_{\nu}\phi-\frac{1}{2}M^{2}\phi^{2}  \notag \\
& & -\frac{1}{2}\kappa h^{\mu\nu}\left[-\partial_{\mu}\phi\partial_{\nu}\phi+
\frac{1}{2}\eta_{\mu\nu}\left(\partial_{\alpha}\phi\partial^{\alpha}
\phi+M^{2}\phi^{2}\right)\right].
\end{eqnarray}

\begin{figure}[t]
\begin{centering}
\includegraphics[scale=1]{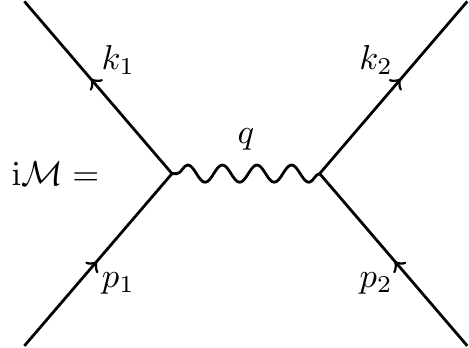}
\par\end{centering}
\caption{The tree-level diagram of two scalar particles interacting via the exchange of a graviton.}
\label{fig:1}
\end{figure}

After this preliminary discussion, we are ready to study the scattering process involving two scalar particles of mass $M_{1}$ and $M_{2}$ exchanging a graviton. The only Feynman diagram contributing to this
process, in lowest order, is drawn in Fig. \ref{fig:1}, and its analytical
expression can be written as
\begin{equation}
i\mathcal{M}=(-i\kappa)^{2}V_{\mu\nu}(p_{1},-k_{1},M_{1})D_{F}^{\mu\nu,\alpha\beta}(q)V_{\alpha\beta}(p_{2},-k_{2},M_{2}),  \label{eq:matrix1}
\end{equation}
where $q=p_{2}-k_{2}=-(p_{1}-k_{1})$ is the momentum transferred, and the vertex
$V^{\mu\nu}(p,k,M)$ corresponds to the expression
\begin{equation}
V^{\mu\nu}(p,k,M)=\frac{1}{2}\left[p^{\mu}k^{\nu}+p^{\nu}k^{\mu}+\eta^{\mu\nu}\left(-p\cdot k+M^{2}\right)\right].  \label{eq:vertex}
\end{equation}
Note that the current conservation is satisfied since $q_{\mu}V^{\mu\nu}=0$ for the external momenta on-shell. Substituting the expressions defined in \eqref{SoftlyProp} and 
\eqref{eq:vertex} into the scattering amplitude \eqref{eq:matrix1}, we
arrive at the following result:
\begin{eqnarray}
i\mathcal{M}	&=&	\frac{i\kappa^{2}}{\frac{\gamma}{2}q^{4}-q^{2}-m^{2}(-q^{2})}\left\{ \frac{\left((D-2)k_{1}\cdot p_{1}+DM_{1}^{2}\right)\left((D-2)k_{2}\cdot p_{2}+DM_{2}^{2}\right)}{4(D-1)}\right.\nonumber\\
		&&-\frac{1}{4}\left[M_{1}^{2}\left((D-2)k_{2}\cdot p_{2}+DM_{2}^{2}\right)+\left(k_{1}\cdot p_{1}\right)\left((D-4)k_{2}\cdot p_{2}+(D-2)M_{2}^{2}\right)\right.\nonumber\\
		&&\left.\left.+2\left(k_{1}\cdot p_{2}\right)\left(k_{2}\cdot p_{1}\right)+2\left(k_{1}\cdot k_{2}\right)\left(p_{1}\cdot p_{2}\right)\right]\right\} 
 \label{matrixLIV}.
\end{eqnarray}

Newtonian dynamics can be accessed by taking the nonrelativistic limit. To that end, we make the approximation (also called static limit) $p_{1,2}=(M_{1,2},0),$ $k_{1,2}=(M_{1,2},0),$ and $q=(0,{\bf q})$. Inserting these expressions into the matrix amplitude \eqref{matrixLIV} and collecting the remaining terms, we get the simplified result
\begin{equation}
i\mathcal{M}_{\mbox{NR}}=-\frac{2 i \kappa^{2}(D-2) M_1^2 M_2^2}{(D-1) \left(\gamma {\bf q}^{4}-2 {\bf q}^{2}-2 m^{2}(-{\bf q}^{2})\right)}.
\label{MatrizNR}\end{equation}

To establish the connection to the Newtonian gravitational potential, we follow Refs. \cite{Accioly:2002tz,Bjerrum-Bohr:2002gqz} and use the Born approximation to the scattering amplitude in nonrelativistic quantum mechanics, defining
\begin{eqnarray}
\left\langle \mbox{f}\left|i\mbox{T}\right|\mbox{i}\right\rangle & \equiv &
(2\pi)^{D}\delta^{D}(p-k)i\mathcal{M}(p_{1},p_{2}\rightarrow k_{1},k_{2})
\notag \\
& \thickapprox & -(2\pi)\delta(E_{p}-E_{k})i\tilde{U}({\bf q}).
\end{eqnarray}
The nonrelativistic energy potential in coordinate space corresponds to
\begin{equation}
U({\bf x}) =  \frac{1}{2M_{1}}\frac{1}{2M_{2}}\int\frac{d^{D-1}q}{(2\pi)^{D-1}
}e^{i{\bf q}\cdot{\bf x}}(-\mathcal{M}_{\mbox{NR}})\label{potCoordinate}.
\end{equation}

From the previous expression of $\mathcal{M}_{\mbox{NR}}$, the Newtonian potential can finally be written in the form
\begin{eqnarray}
U({\bf x}) &=&\frac{\kappa^{2}M_{1}M_{2}}{2(2\pi)^{D-1}}\frac{(D-2)}{(D-1)} \int_{0}^{\infty}\int_{0}^{\pi}\cdots\int_{0}^{\pi}\int_{0}^{2\pi}\left[\frac{e^{i{\bf q}\cdot{\bf x}}}{\gamma {\bf q}^{4}-2 {\bf q}^{2}-2 L^{2(\alpha-1)}({\bf q}^{2})^{\alpha}}|{\bf q}|^{D-2}d|{\bf q}|\right.\nonumber\\
&&\left.\times\sin^{D-3}\theta_{D-2}d\theta_{D-2}\cdots\sin^{2}\theta_{3}d\theta_{3}\sin\theta_{2}d\theta_{2}d\theta_{1}\right] \label{potCoordinate2},
\end{eqnarray}
where we have substituted $m^{2}(-{\bf q}^{2})=L^{2(\alpha-1)}({\bf q}^{2})^{\alpha}$.

Consequently, the problem of calculating the effective nonrelativistic potential was reduced to solving Fourier integrals. In what follows, we will apply our result in two examples to illustrate the efficiency of the method: we study linearized softly massive quadratic gravity in three and four spacetime dimensions, respectively.

%%%%%%%%%%%%%%%%%%%%%%%%%%%%%%%%%%%%%%%%%%%%%%%%%%%%%%%%%%%%%%%%%%%%%%%

\subsection{Three-dimensional linearized softly massive quadratic gravity}

For the three-dimensional DGP case $(\gamma=0,\alpha=1/2)$, we have the potential
\begin{eqnarray}
U(r) &=&-\frac{G M_1 M_2}{\pi}\int_{0}^{\infty}|{\bf q}|d|{\bf q}|\int_{0}^{2\pi}d\theta\frac{e^{i|{\bf q}| r \cos\theta}}{{\bf q}^2+m|{\bf q}|}\nonumber\\
&=&-G M_{1} M_{2}\left[-2{}_0F_{1REG}^{(1,0)}\left(1,-\frac{1}{4} m^2 r^2\right)\right.\nonumber\\
&&+\left.\pi\pmb{H}_0(m r)+2\ln\left(\frac{2}{mr}\right)J_0(m r)\right],
\end{eqnarray}where $J_{n}(z)$ is the Bessel function of the first kind, $\pmb{H}_n(z)$ is the Struve function and ${}_0F_{1REG}^{(1,0)}(a;z)$ representing a function obtained from the regularized confluent hypergeometric function  $_{0}F_{1}(a;z)/ \Gamma(a)$ by one differentiating with respect to the first argument. The coupling constant $G$ is the effective gravitational
one defined as $G\equiv \kappa^{2}/32\pi$. 

Two relevant situations are, of course, the short and long distance behavior of the potential.  First, let us introduce the characteristic 
distance scale: 
\begin{equation}
r_{0}\equiv \frac{1}{m}.
\end{equation}
At short distances when $r\ll r_{0}$, the potential becomes
\begin{eqnarray}
U(r) & \simeq &-2G M_{1}M_{2}\left[\ln\left(\frac{r_{0}}{r}\right)+\frac{r}{r_{0}}+\ln2-{}_{0}F_{1REG}^{(1,0)}\left(1,0\right)\right]+\mathcal{O}(r^{2})
	\sim  \ln r,%\nonumber
\end{eqnarray}
which has a logarithmic singularity at the origin, usual characteristic of the $3D$ Newtonian potential. At the opposite extreme, when $r \gg r_{0}$, the potential gives
\begin{eqnarray}
U(r) & \simeq &-2GM_{1}M_{2}\frac{r_{0}}{r}+\mathcal{O}(\frac{1}{r^{2}})
 \sim  -\frac{1}{r},%\nonumber
\end{eqnarray}
which agrees with Newton's law in $4D$. These results show a peculiar feature of the DGP theory; it interpolates the dimensional behavior of gravity between $D$ at short distance and $D+1$ at large distance \cite{Hao:2014tsa}.

%%%%%%%%%%%%%%%%%%%%%%%%%%%%%%%%%%%%%%%%%%%%%%%%%%%%%%%%%%%%%%%%
For the higher-derivative DGP case $(\gamma\neq 0,\alpha=1/2)$, we have the effective potential
\begin{eqnarray}
\label{potential}
U(r) &=&\frac{2 G M_1 M_2}{\pi}\int_{0}^{\infty}|{\bf q}|d|{\bf q}|\int_{0}^{2\pi}d\theta\frac{e^{i|{\bf q}| r \cos\theta}}{\gamma {\bf q}^{4}-2{\bf q}^2-2m|{\bf q}|}\nonumber\\
&=& 4 G M_1 M_2\int_{0}^{\infty}\frac{J_0(q r)}{\gamma  q^3-2 (q+m)}dq.
\end{eqnarray}

\begin{figure}[th]
\begin{centering}
\includegraphics[scale=0.5]{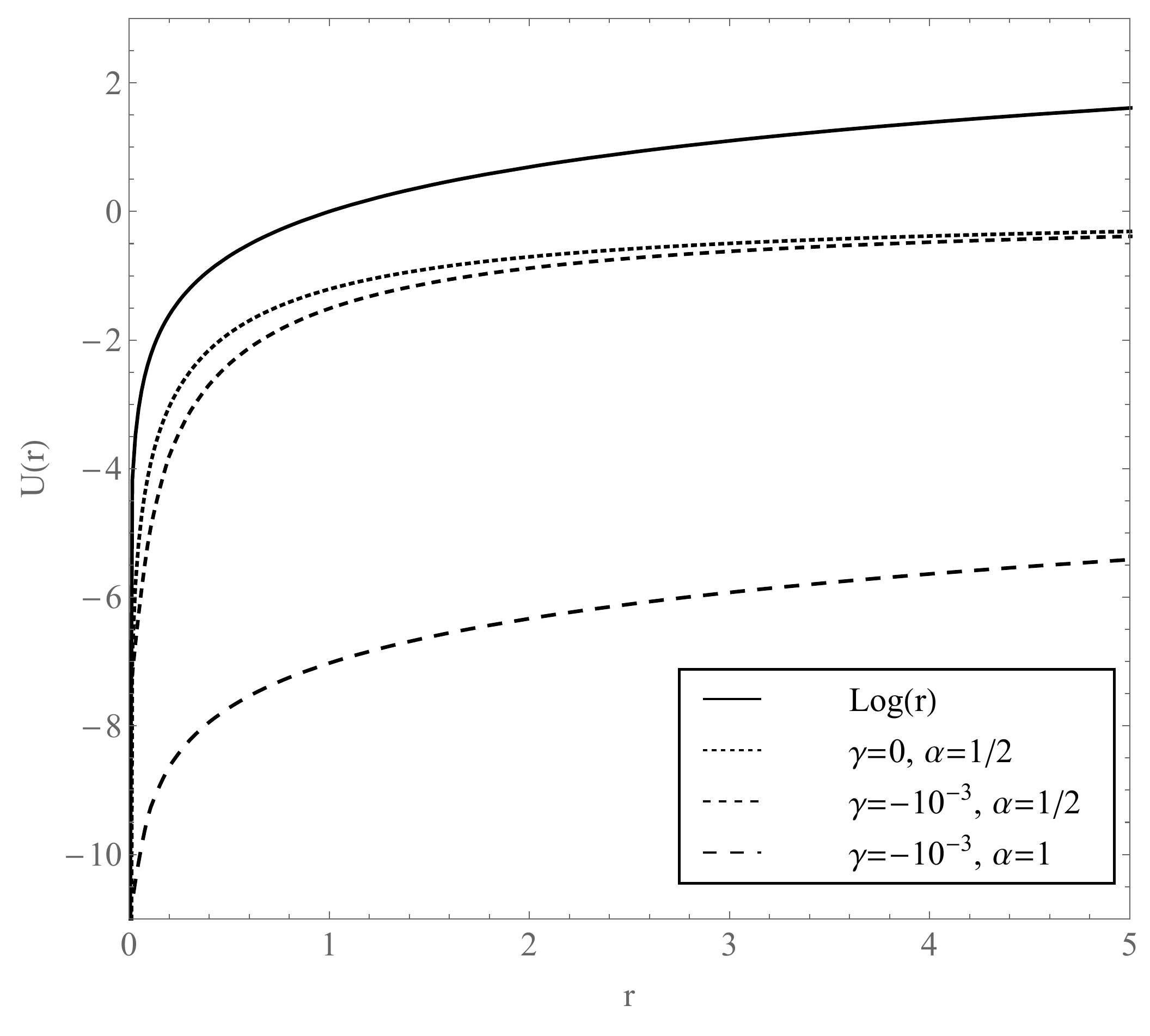}
\par\end{centering}
\caption{The effective potential for $3D$ linearized softly massive quadratic gravity.}
\label{fig:2}
\end{figure}

The last integral cannot be solved in terms of any standard mathematical functions. But it is convergent for $\gamma<0$, and we plot it in Fig. \ref{fig:2}.

Suggesting that $m$ is small, we can do some estimation for this integral: at $q<m$, we approximate the integrand by its low-argument asymptotics, i.e. $J_0(qr)\simeq 1$, and at $q>m$ -- by high-argument asymptotics, i.e.  the denominator is approximated by $\gamma q^3$, and $J_0(qr)\simeq \sqrt{\frac{2}{\pi qr}}\cos qr$. So, at $\gamma<0$, we arrive at
\begin{eqnarray}
U(r) &=& -4 G M_1 M_2
\Big(
\int_0^m\frac{dq}{2 (q+m)}+\frac{1}{|\gamma|}\int_m^{\infty}\frac{dq}{q^3}\sqrt{\frac{2}{\pi qr}}\cos qr
\Big).
\end{eqnarray}
The first term can be integrated immediately, and the second one represents itself as an integral from a rapidly oscillating function, so, we can estimate this integral as
\begin{eqnarray}
U(r) &=& -4 G M_1 M_2
\Big(
\frac{1}{2}\ln 2+\frac{1}{|\gamma|}\frac{1}{m^2}\sqrt{\frac{2}{\pi mr}}\cos mr
\Big).
\end{eqnarray}
For the higher-derivative softly massive case $(\gamma\neq 0,\alpha=1)$, we have the energy potential
\begin{eqnarray}
\label{epo}
U(r) &=&\frac{2 G M_1 M_2}{\pi}\int_{0}^{\infty}|{\bf q}|d|{\bf q}|\int_{0}^{2\pi}d\theta\frac{e^{i|{\bf q}| r \cos\theta}}{\gamma {\bf q}^{4}-4{\bf q}^2}\nonumber\\
&=& 4 G M_1 M_2\int_{0}^{\infty}\frac{J_0(q r)}{\gamma  q^3-4q}dq\nonumber\\
&=& 4 G M_1 M_2\int_{0}^{\infty}\left[\frac{\gamma  q}{4 \left(\gamma  q^2-4\right)}-\frac{1}{4 q}\right]J_0(q r)dq\nonumber\\
&=&G M_1 M_2 K_0\left(\frac{2 r}{\sqrt{-\gamma }}\right)-G M_1 M_2 \int_{0}^{\infty}\frac{J_0(q r)}{q}dq,
\end{eqnarray}
with the last integral being infrared divergent $(q\rightarrow 0)$. Indeed, we can verify that
\begin{equation}
\lim_{\epsilon\to 0^+}\left[\ln\epsilon+\int_{\epsilon}^\infty dk\frac{J_0(kr)}{k}\right]=\ln \left(\frac{2}{r}\right)-\gamma_{E},
\end{equation}
where  $\gamma_{E}\approx0.577\cdots$ is the Euler's constant. 
We note that this potential indeed displays a logarithmic behavior at small distances as it must be in three dimensions since $K_0(x)\propto  \ln x$. At large distances, the first term of (\ref{epo}) is suppressed decaying exponentially. In Fig. \ref{fig:2} we plot the profile of the three-dimensional potential $U(r)$, for some values of $\gamma$ and $\alpha$.

\subsection{Four-dimensional linearized softly massive quadratic gravity}

For the standard DGP case $(\gamma=0,\alpha=1/2)$, we have the energy potential
\begin{eqnarray}
U(r) &=&-\frac{2 G_{N} M_1 M_2}{3 \pi ^2}\int_{0}^{\infty}{\bf q}^{2}d|{\bf q}|\int_{0}^{\pi}\sin\theta d\theta\int_{0}^{2\pi}d\phi\frac{e^{i|{\bf q}| r \cos\theta}}{{\bf q}^2+m|{\bf q}|}\nonumber\\
&=&-\frac{8 G_{N} M_1 M_2 }{3 \pi  r}\left\{ \sin\left(\frac{r}{r_{0}}\right)\text{Ci}\left(\frac{r}{r_{0}}\right)+\frac{1}{2}\cos\left(\frac{r}{r_{0}}\right)\left[\pi -2\,\text{Si}\left(\frac{r}{r_{0}}\right)\right]\right\},
\end{eqnarray}
where 
\begin{eqnarray}
\text{Si}(x)&\equiv&\int_{0}^{x}\frac{dt}{x}\sin t,\nonumber\\
\text{Ci}(x)&\equiv&\gamma_{E}+\ln x+\int_{0}^{x}\frac{dt}{t}(\cos t-1).
\end{eqnarray}
Note that $U(r)$ behaves as
\begin{equation}
U(r)=\left\{ \begin{array}{cc}
-\frac{4G_{N}M_{1}M_{2}}{3}\frac{1}{r}-\frac{8G_{N}M_{1}M_{2}}{3\pi}\frac{1}{r_{0}}\left[\gamma_{E}-1+\ln\left(\frac{r}{r_{0}}\right)\right]+\mathcal{O}(r), & r\ll r_{0},\\
-\frac{8G_{N}M_{1}M_{2}}{3\pi}\frac{r_{0}}{r^{2}}+\mathcal{O}\left(\frac{1}{r^{3}}\right), & r\gg r_{0}.
\end{array}\right.
\end{equation}and, as expected, the potential interpolates between the standard $4D$ asymptotics $\sim 1/r$ and $5D$ one $\sim 1/r^{2}$ at the length scale defined by $r_{0}$ \cite{Hinterbichler:2011tt}.

For the higher-derivative DGP case $(\gamma\neq 0,\alpha=1/2)$, we have the energy potential
\begin{eqnarray}
U(r) &=&\frac{4 G_{N} M_1 M_2}{3 \pi ^2}\int_{0}^{\infty}{\bf q}^{2}d|{\bf q}|\int_{0}^{\pi}\sin\theta d\theta\int_{0}^{2\pi}d\phi\frac{e^{i|{\bf q}| r \cos\theta}}{\gamma {\bf q}^4-2{\bf q}^2-2 m|{\bf q}|}\nonumber\\
&=&\frac{16 G_{N} M_1 M_2}{3\pi r}\int_{0}^{\infty}\frac{\sin (r |{\bf q}| )}{\gamma {\bf q}^3-2(|{\bf q}|+m)}d|{\bf q}|.
\end{eqnarray}
We approximate this integral again by its low-argument asymptotics at small ${\bf q}$, and higher-argument asymptotics at large ${\bf q}$ as above, see discussion after (\ref{potential}). Again, for $\gamma<0$ we have $U({\bf x}) =-\frac{16 G m_1 m_2}{3\pi r} I$, where
$$
I=\int_{0}^{\infty}\frac{\sin (r |{\bf q}| )}{|\gamma| {\bf q}^3+2(|{\bf q}|+m)}d{\bf q}\simeq \frac{1}{2m}
(1-\cos mr)+\frac{\sin mr}{2|\gamma|m^3}.$$
We see that taking the integral $I$ into account, one finds, besides of usual $r^{-1}$ terms, also oscillating terms which also decay as distance grows.
 %This last integral now involves an implicit sum over the roots of a polynomial.

\begin{figure}[th]
\begin{centering}
\includegraphics[scale=0.5]{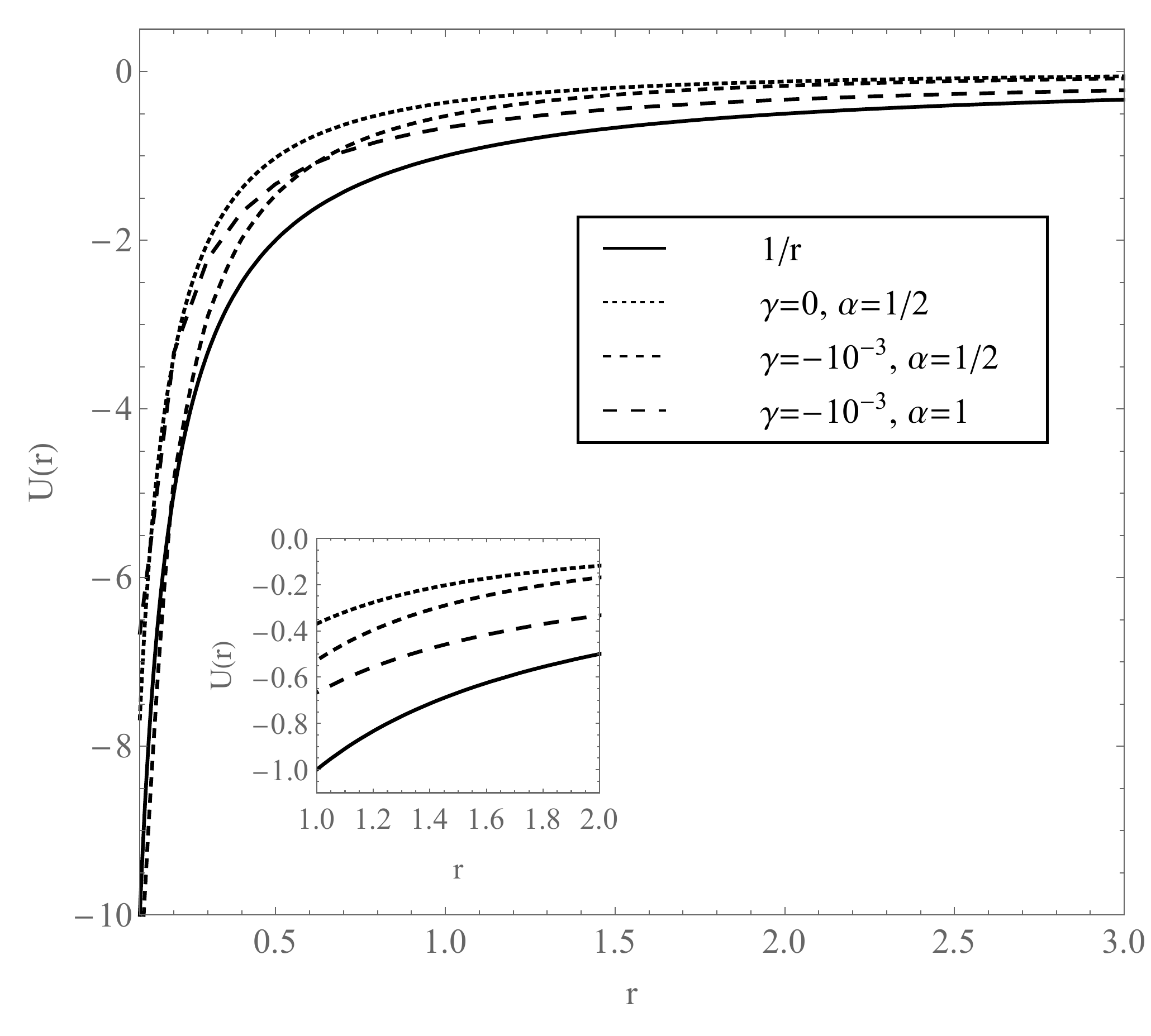}
\par\end{centering}
\caption{The effective potential for $4D$ linearized softly massive quadratic gravity.}
\label{fig:3}
\end{figure}

For the higher-derivative softly massive case $(\gamma\neq 0,\alpha=1)$, we have the energy potential
\begin{eqnarray}
U(r) &=&\frac{4 G_{N} M_1 M_2}{3 \pi ^2}\int_{0}^{\infty}{\bf q}^{2}d|{\bf q}|\int_{0}^{\pi}\sin\theta d\theta\int_{0}^{2\pi}d\phi\frac{e^{i|{\bf q}| r \cos\theta}}{\gamma {\bf q}^4-4{\bf q}^2}\nonumber\\
&=&-\frac{4 G_{N} M_1 M_2 }{3r}\exp\left(-\frac{r}{\sqrt{|\gamma|}}\right) \sinh\left(\frac{r}{\sqrt{|\gamma|}}\right),
\end{eqnarray}
where we assume $\gamma<0$ again. It is interesting to note that for large arguments, the potential tends to a constant. Fig. (\ref{fig:3}) shows the profile of the four-dimensional potential $U(r)$, for some values of $\gamma$ and $\alpha$.

%
%
%%%%%%%%%%%%%%%%%%%%%%%%%%%%%%%%%%%%%%%%%%%%%%%%%%%%%%%%%%%%%%%%%%%%%%%%%%%%%%%%%
\section{Softly massive gravity with a gravitational Chern-Simons and Ricci-Cotton terms}
\label{sec4}

Next, in this section, we will consider the more general situation in which different mass terms can coexist in the context of the three-dimensional softly massive gravity. Specifically, we are interested in studying the Pauli-Fierz-like soft mass term with an additive gravitational Chern-Simons one. In particular, there is an additional higher-derivative topological term named the gravitational Ricci-Cotton term in three dimensions. So, a natural question is what happens to the graviton propagator if these various terms coexist in a single theory. As we will see below, these apparently innocuous models are not physically acceptable due to the appearance of ghosts and/or tachyons in the graviton spectrum.

The general action including all the contributions mentioned above can be written as
\begin{equation}
S=\int d^3 x\left[\sqrt{-g}\frac{2}{\kappa^{2}}\left\{R+\frac{\beta}{2}R^{2}+\frac{\gamma}{2}R^{2}_{\mu\nu}\right\}-\frac{1}{2}\left(h_{\mu\nu}m^{2}(\Box)h^{\mu\nu}-h m^{2}(\Box) h \right)+\mathcal{L}_{CS}+\mathcal{L}_{RC}\right],\label{action3D}
\end{equation}where the topological Chern-Simons term is given by
\begin{equation}
\mathcal{L}_{CS}=\frac{\mu}{2}\epsilon^{\mu\nu\rho}\Gamma^{\sigma}_{\mu\lambda}\left(\partial_{\nu}\Gamma^{\lambda}_{\sigma\rho}+\frac{2}{3}\Gamma^{\lambda}_{\nu\omega}\Gamma^{\omega}_{\rho\sigma}\right),
\end{equation}where $\mu$ is a dimensionless parameter, and the Ricci-Cotton term is defined by the Lagrangian density
\begin{equation}
\mathcal{L}_{RC}=\lambda \epsilon^{\mu\nu\rho}R_{\mu\sigma}D_{\nu}R^{\ \ \sigma}_{\rho},
\end{equation}
where $\lambda$ has dimension of mass. Now, it is worth mentioning that the topological nature of the CS term is related to the four-dimensional Chern-Pontryagin topological density, $\mathcal{P}_{4}$, which in turn can be written as a total derivative of a four-dimensional vector, i.e.,
\begin{equation}
\mathcal{P}_{4}=\partial_{m}J^{m},
\end{equation} 
where 
\begin{equation}
J^{m}=\epsilon^{mnop}\Gamma^{q}_{nl}\left(\partial_{o}\Gamma^{l}_{pq}+\frac{2}{3}
\Gamma^{l}_{or}\Gamma^{r}_{pq}\right)
\end{equation}
is the Chern-Simons topological current. Having said this, we note that $\mathcal{L}_{CS}$ is just proportional to a particular component
of $J^{m}$, say, $J^{3}$. Then, $\mathcal{L}_{CS}\propto J^{3}$. Using similar arguments, one can show that the Ricci-Cotton term may be rewritten as a total derivative of another topological current. Apart from that, despite the topological character of CS and RC terms, they are defined by construction in terms of the Christoffel symbols which are entirely described by the metric. Consequently, such terms dramatically affect the gravitational field equations -- for instance -- we invite the reader to the work \cite{Myung:2009sk} to see how the CS term contributes to the bulk field equations.

 In order to accommodate these topological terms, we have to introduce two additional operators to the whole set of spin projection operators \cite{Nakasone:2009vt} 
\begin{eqnarray}
S_{1\mu\nu,\rho\sigma}&=&\frac{1}{4}\Box(\epsilon_{\mu\rho\lambda}\partial_\sigma\omega^{\lambda}_{\nu}+\epsilon_{\mu\sigma\lambda}
\partial_{\rho}\omega^{\lambda}_{\nu}+\epsilon_{\nu\rho\lambda}\partial_{\sigma}\omega^{\lambda}_{\mu}
+\epsilon_{\nu\sigma\lambda}\partial_{\rho}\omega^{\lambda}_{\mu}),\label{Sp1}\\
S_{2\mu\nu,\rho\sigma}&=&-\frac{1}{4}\Box(\epsilon_{\mu\rho\lambda}\eta_{\sigma\nu}+\epsilon_{\mu\sigma\lambda}\eta_{\rho\nu}
+\epsilon_{\nu\rho\lambda}\eta_{\sigma\mu}+\epsilon_{\nu\sigma\lambda}\eta_{\rho\mu})\partial^{\lambda}.\label{Sp2}
\end{eqnarray}
These operators obey the following tensorial relations involving the other spin projection operators:
\begin{eqnarray}
&& S_{1}S_{1}=\frac{1}{4}\Box^{3}P^{(1)},\ \ \ \ S_{1}S_{2}=S_{2}S_{1}=-\frac{1}{4}\Box^{3}P^{(1)},\nonumber\\
&& S_{2}S_{2}=\Box^{3}\left(P^{(2)}+\frac{1}{4}P^{(1)}\right),\ \ \ \ P^{(1)}S_{1}=S_{1}P^{(1)}=S_{1},\nonumber\\
&& P^{(1)}S_{2}=S_{2}P^{(1)}=-S_{1},\ \ \ \ P^{(2)}S_{2}=S_{2}P^{(2)}=S_{1}+S_{2},\label{Sprojector2}
\end{eqnarray}
where the tensor indices are suppressed for the sake of simplicity.

Following the same methodology described earlier, now aided by relations (\ref{Sp1}) and (\ref{Sp2}), we can collect the quadratic fluctuations in $h_{\mu\nu}$ at the action (\ref{action3D}), and the result reads
\begin{equation}
S=\int d^{3}x\frac{1}{2}h^{\mu\nu}\mathcal{Q}_{\mu\nu,\alpha\beta} h^{\alpha\beta},
\end{equation} where $\mathcal{Q}_{\mu\nu,\alpha\beta}$ is defined as
\begin{eqnarray}
\mathcal{Q}_{\mu\nu,\alpha\beta} &=&\left[\frac{\gamma}{2}\Box^{2}+\Box -m^{2}(\Box)\right]P^{(2)}_{\mu\nu,\alpha\beta}-m^{2}(\Box) P^{(1)}_{\mu\nu,\alpha\beta}\nonumber\\
&+&\left[\left(\frac{3}{2}\gamma + 4\beta\right)\Box^{2}-\Box + m^{2}(\Box)\right]P^{(0,s)}_{\mu\nu,\alpha\beta}\nonumber\\
&+&\sqrt{2}m^{2}(\Box)(P^{(0,sw)}+P^{(0,ws)})_{\mu\nu,\alpha\beta}\nonumber\\
&+&\left(\frac{\mu}{2}+\frac{\lambda}{2}\Box\right)(S_1+S_2)_{\mu\nu,\alpha\beta}.
\end{eqnarray}

Accordingly, with the help of the multiplication relations (\ref{orthog}) and (\ref{Sprojector2}), and taking into account that $\mathcal{Q}\mathcal{Q}^{-1}=\mathcal{I}$, we find  that the inverse matrix of $\mathcal{Q}_{\mu\nu,\alpha\beta}$ is given by 
\begin{eqnarray}
\mathcal{Q}^{-1}_{\mu\nu,\alpha\beta}	&=&	\left[\frac{\frac{\gamma}{2}\Box^{2}+\Box-m^{2}(\Box)}{\left(\frac{\gamma}{2}\Box^{2}+\Box-m^{2}(\Box)\right)^{2}-\frac{1}{4}\Box^{3}(\mu+\lambda\Box)^{2}}\right]P^{(2)}_{\mu\nu,\alpha\beta}-\left[\frac{1}{m^{2}(\Box)}\right]P^{(1)}_{\mu\nu,\alpha\beta}\nonumber\\
&+&\left[\frac{(-8\beta-3\gamma)\Box^{2}+2\Box-2 m^{2}(\Box)}{4(m^{2}(\Box))^{2}}\right]P^{(0,w)}_{\mu\nu,\alpha\beta}+\left[\frac{1}{\sqrt{2}m^{2}(\Box)}\right](P^{(0,sw)}+P^{(0,ws)})_{\mu\nu,\alpha\beta}\nonumber\\
&+&\left[\frac{\lambda\Box+\mu}{2\left(\frac{1}{4}\Box^{3}(\lambda\Box+\mu)^{2}-\left(\frac{\gamma}{2}\Box^{2}+\Box-m^{2}(\Box)\right)^{2}\right)}\right](S_1+S_2)_{\mu\nu,\alpha\beta}.
\end{eqnarray}
Note that this expression reduces to the one obtained in (\ref{invO}) for $D=3$ when $\lambda,\mu\rightarrow 0$, as expected.

Differently to the previous case without the topological Chern-Simons and Ricci-Cotton terms, there appears a quartic pole in the sectors of $P^{(2)}$, $S_1$, and $S_2$. In general, higher order poles are indicative of ghosts and/or tachyons in a given theory. The explicit analysis of the presence of such instabilities can be done in much the same way to the one performed in the end of Section (\ref{sec2}). By proceeding in this way, we conclude that the possible massive poles can be read off from the function $J\equiv \left(\frac{\gamma}{2}\Box^{2}+\Box-m^{2}(\Box)\right)^{2}-\frac{1}{4}\Box^{3}(\mu+\lambda\Box)^{2}$ . It can be verified that the equation  $J=0$ has real solutions for several possible choices of parameters. Hence, the presence of ghosts and tachyons in the spectrum of the theory is possible.

\section{Final Remarks}
\label{conclu}
We considered a toy model for the DGP braneworld description for a 
massive higher-derivative gravity theory in $D$ dimensions, with the softly nonlocal mass term emerges due to the presence of the extra dimension within the DGP framework. As a result, we succeeded to obtain corrections to the Newtonian potential explained by the presence of a nontrivial mass of the graviton. We calculated the propagator and found that ghosts and tachyons can arise in the spectrum of the theory, depending on certain choices of the free parameters present in the model. So, we have a situation similar to \cite{Nakasone:2009vt,Gregory:2000jc} where the tachyons and ghosts can be present. Then, we found the analogue of the Newtonian potential.  We demonstrated that in various cases, besides of the usual logarithmic-like behavior, corrections to the potential behaving like constants or decaying with a distance are possible.

We expect that these conclusions can be generalized to the usual case of a four-dimensional brane immersed into a five-dimensional space. We suggest to carry this study in a forthcoming paper.

%%%%%%%%%%%%%%%%%%%%%%%%%%%%%%%%%%%%%%%%%%%%%%%%%%%%%%%%%%%%%%%%%%%%%%%%%%%%%%%%%%%%%
\begin{acknowledgments}

The authors would like to thank the Funda\c{c}\~{a}o Cearense de Apoio ao Desenvolvimento
Cient\'{\i}fico e Tecnol\'{o}gico (FUNCAP), the Coordena\c{c}\~ao de Aperfei\c{c}oamento de Pessoal de N\'ivel Superior (CAPES), and the Conselho Nacional de Desenvolvimento Cient\'{\i}fico e Tecnol\'{o}gico (CNPq) for
financial support. The work by A. Yu. P. has been supported by the
CNPq project No. 301562/2019-9. P. J. P. would like to thank the Brazilian agency CAPES for financial support (PNPD/CAPES grant, process 88887.464556/2019-00).

\end{acknowledgments}

%%%%%%%%%%%%%%%%%%%%%%%%%%%%%%%%%%%%%%%%%%%%%%%%%%%%%%%%%%%%%%%%%%%%%%%%%%%%%%%%%%%%%

\end{document}